\documentclass[twocolumn,prl,amsmath,amssymb]{revtex4}
\usepackage{graphicx}
\usepackage{epstopdf}
\usepackage{amsmath}
\DeclareGraphicsExtensions{.pdf,.eps,.png,.jpg,.mps}

\begin{document}
\title{Doping dependence of the Raman spectrum of defected graphene}
\author{M. Bruna, A. K. Ott, M. Ij\"as, D. Yoon, U. Sassi, A. C. Ferrari}
\email{acf26@eng.cam.ac.uk}
\affiliation{Cambridge Graphene Centre, University of Cambridge, Cambridge CB3 0FA, UK}
\begin{abstract}
We investigate the evolution of the Raman spectrum of defected graphene as a function of doping. Polymer electrolyte gating allows us to move the Fermi level up to 0.7eV, as monitored by \textit{in-situ} Hall-effect measurements. For a given number of defects, the intensities of the D and D' peaks decrease with doping. We assign this to an increased total scattering rate of the photoexcited electrons and holes, due to the doping-dependent strength of electron-electron scattering. We present a general relation between D peak intensity and defects valid for any doping level.
\end{abstract}
\maketitle
\section{\label{In}Introduction}
Raman spectroscopy is one of the most used characterization techniques in carbon science and technology\cite{acftrans}. The measurement of the Raman spectrum of graphene\cite{Ferrari2006} triggered a huge effort to understand phonons, electron-phonon, magneto-phonon and electron-electron interactions in graphene, as well as the influence of the number and orientation of layers, electric or magnetic fields, strain, doping, disorder, quality and types of edges, and functional groups\cite{Ferrari_NNano2013}.

Quantifying defects in graphene is crucial both to gain insight in fundamental properties, and for applications. Ref.\citenum{Ferrari2000} introduced a three-stage classification of disorder, leading from graphite to amorphous carbons, that allows to simply assess all the Raman spectra of carbons: stage 1) graphene to nanocrystalline graphene; stage 2) nanocrystalline graphene to low-$sp^3$ amorphous carbon; stage 3) low-$sp^3$ amorphous carbon to high-$sp^3$ amorphous carbon. Here we focus on stage 1, the most relevant when considering the vast majority of publications dealing with graphene production, processing and applications. In stage 1 the Raman spectrum evolves as follows\cite{Ferrari2000}: a) D appears and the ratio of the D and G peak intensities, I(D)/I(G), increases; b) D' appears; c) all peaks broaden; e) the D+D' peak appears; f) at the end of stage 1, G and D' are so wide that is sometimes more convenient to consider them as a single, up-shifted, wide G band at$\sim1600\:\mbox{cm}^{-1}$.

In their seminal work, Tuinstra and Koenig noted that I(D)/I(G) varied inversely with the crystal size, $L_a$: I(D)/I(G)=$C(\lambda)/L_a$, where $C(514\:\mbox{nm})\sim{4.4}\:\mbox{nm}$\cite{TuinstraKoenig,Matthews,Knight} ($\lambda$ being the excitation wavelength). Initially, this was interpreted in terms of phonon confinement: the intensity of the forbidden process would be ruled by the ``amount of lifting'' of the selection rule\cite{TuinstraKoenig}, $\Delta q\propto 1/\Delta x$, with $\Delta x\approx L_a$. Now, it is understood theoretically and established experimentally, that the D peak is produced only in a small region of the crystal (size $\sim{v}_F/\omega_D\sim{3-4}\:\mbox{nm}$, where $\sim{v}_F$ is the Fermi velocity and $\omega_D$ is the phonon frequency) near a defect or an edge\cite{Casiraghi_edges2009,lucchese,Beams}. For a nanocrystallite, I(G) is proportional to the sample area,$\propto{L}_a^2$, while I(D) is proportional to the overall length of the edge, which scales as$\sim L_a$. Thus, I(D)/I(G)$\propto 1/L_a$. For a sample with rare defects, I(D) is proportional to the total number of defects probed by the laser spot. Thus, for an average interdefect distance $L_D$, and laser spot size $L_L$, there are on average $(L_L/L_D)^2$ defects in the area probed by the laser, then I(D)$\propto(L_L/L_D)^2$. On the other hand, I(G) is proportional to the total area probed by the laser $\propto(L_L)^2$, thus I(D)/I(G)=$C''(\lambda)/L_D^2$. For very small $L_D$, one must have $C''(\lambda)/L_D^2=$I(D)/I(G)$=C(\lambda)/L_a$. This condition gives an estimate of $C''(514\:\mbox{nm})\sim{90}\:\mbox{nm}$. Ref.\citenum{lucchese} measured I(D)/I(G) for irradiated single layer graphene (SLG) with known $L_D$, derived from STM measurements, obtaining I(D)/I(G)$\approx 145/L_D^2$ at 514~nm excitation, in excellent agreement with this simple estimate.

Ref.\citenum{Cancado2011} then considered the excitation energy dependence of the peaks areas and intensities, for visible excitation energy. A fit to the experimental data gave the relation\cite{Cancado2011}:
\begin{equation}\label{eq12}
L_D^2\,{\rm(nm^{2})}=\frac{4.3\times10^{3}}{E_L^4(\mbox{eV}^4)}\left[\frac{\mathrm{I(D)}}{\mathrm{I(G)}}\right]^{-1}
\end{equation}
where $E_L$ is the laser excitation energy.

By considering point-like defects, separated from each other by $L_D$[nm], Eq.1 can be restated in terms of defect density $n_{\rm D}$(cm$^{-2}$)\,=\,10$^{14}$/$[\pi L_{\rm D}^{2}(nm^2))]$\cite{Cancado2011}:
\begin{equation}\label{eq14}
n_{\rm D}({\rm cm}^{-2})=7.3\times10^{9}E_L^{4}(\mbox{eV}^4)\,\frac{\mathrm{I(D)}}{\mathrm{I(G)}}
\end{equation}
Note that these relations are limited to Raman-active defects. Perfect zigzag edges\cite{Cancado2004,Casiraghi_edges2009}, charged impurities\cite{Casiraghi07_ChargedImp,Das2008}, intercalants\cite{tandoping}, uniaxial and biaxial strain\cite{Mohiuddin2009,Halsall} do not generate a D peak. For these types of ``silent'' defects, other Raman signatures can be used. A perfect edge does change the G peak shape\cite{Sasaki09,Cong10}, while strain, intercalants, and charged impurities have a strong influence on the G and 2D peaks\cite{Casiraghi07_ChargedImp,Das2008,tandoping,Mohiuddin2009}. In this case, the combination of Raman spectroscopy with other independent probes of the number of defects can provide a wealth of information on the nature of such defects.

We note as well that these relations are derived assuming negligible Fermi level, $E_F$, shift. It is known that doping has major effects on the Raman spectra\cite{Pisana2007,Das2008,Basko2009ee}. The G peak position, Pos(G), increases and its Full Width at Half Maximum, FWHM(G), decreases for both electron ($e$) and hole ($h$) doping. The G peak stiffening is due to the non-adiabatic removal of the Kohn anomaly at the Brillouin Zone (BZ) centre, $\Gamma$\cite{Pisana2007}. The FWHM(G) sharpening is due to Pauli blocking of phonon decay into $e-h$ pairs when the $e-h$ gap is higher than the phonon energy\cite{LazPRB2006,Pisana2007}, and saturates for $E_F$ bigger than half phonon energy\cite{Pisana2007,LazPRB2006}. Furthermore, in SLG the ratio of the heights of the 2D and G peaks, I(2D)/I(G), and their areas, A(2D)/A(G), is maximum for zero doping\cite{Ferrari2006,berciaud}, and decreases for increasing doping. The doping dependence of the 2D intensity results from its sensitivity to the scattering of the photoexcited $e$ and $h$. Assuming the dominant sources of scattering to be phonon emission and $e-e$ collisions, Ref.\citenum{Basko2009ee} showed that, while the former is not sensitive to doping, the latter is. Then, the doping dependence of the 2D peak can be used to estimate the corresponding electron-phonon coupling\cite{Basko2009ee}. These considerations apply for $|E_F|$ small compared to $\hbar\omega_L/2$ ($\omega_L$ being the angular frequency of the incident photon). In the past few years, much higher doping levels have been achieved\cite{Kalbac2010,Chen2011,tandoping}. One of the effects of high doping is the increase in I(G). Doping changes the occupations of electronic states and, since transitions from an empty state or to a filled state are impossible, it can exclude some BZ regions from contributing to the Raman matrix element. Due to suppression of destructive interference, this leads to an enhancement of the G peak when $|E_F|$ matches $\hbar\omega_L/2$, as predicted theoretically\cite{BaskoNJP} and observed experimentally\cite{Kalbac2010,Chen2011}. Another effect of high doping is on the 2D peak, which is suppressed when the conduction band becomes filled at the energy probed by the laser\cite{tandoping}. There are three cases: (i) when $\omega_L,\omega_{Sc}>2|E_F|/\hbar$ (where $\omega_{Sc}$ is the angular frequency of the emitted photon), all processes are allowed and the 2D band is observed, (ii) when $\omega_{Sc}<2|E_F|/\hbar<\omega_L$, the photon absorption is allowed but the phonon emission is excluded by Pauli blocking; (iii) when $\omega_L,\omega_{Sc}<2|E_F|/\hbar$, both photon absorption and phonon emission are blocked. Therefore, only when $2|E_F|/\hbar<[\omega_L-\mbox{Pos}(2\mathrm{D})$], the 2D band is observable.

While a significant effort was devoted to understand the effect of defects in samples with negligible doping\cite{Ferrari2000, Casiraghi_edges2009, Cancado2011}, and the effect of doping in samples with negligible defects\cite{Das2008, Basko2009ee, BaskoNJP, tandoping, Kalbac2010,Chen2011}, the combined effect of doping and defects on the Raman spectrum of SLG has received little attention. However, most samples produced by either micromechanical exfoliation, chemical vapor deposition, liquid phase exfoliation or carbon segregation from SiC or metal substrates are naturally doped due to the extreme sensitivity of graphene to the presence of adsorbates (e.g. moisture) and to the interaction with the underlying substrate\cite{Casiraghi07_ChargedImp,Das2008,Bonaccorso2012Production}. Additionally, many of them also have defects, or defects may appear during processing for device integration. It is thus critical to understand if and how defects can be detected and quantified by Raman spectroscopy in doped samples.

Here we study the dependence of the Raman spectrum of defected SLG on the level of electrostatic doping, in samples with a fixed amount of defects. We combine polymer electrolyte gating\cite{Das2008} with \textit{in situ} Hall-effect measurements and Raman spectroscopy at different excitation wavelengths. This allows us to vary $E_F$ from $\approx -0.7$eV ($h-$doping) up to $\approx 0.4$eV ($e-$doping), a much wider span than what can be achieved by the common 300nm SiO$_2$back gate (usually restricted to $\approx \pm 0.3eV$ because of the limited gate capacitance\cite{GeimNatMat07}) and large enough to cover the range of doping found in the vast majority of papers in literature. This range is however smaller than $\hbar\omega_L/2$ to exclude additional effects due to Pauli blocking on the peaks' intensities, again consistent with the case in most papers. We find that the intensity of defect-related peaks, D and D', strongly decreases with $E_F$. We assign this to increased broadening of the electronic states due to increased $e-e$ scattering for higher doping. We then modify Eqs.(1,\ref{eq14}) and give general relations between D peak intensity and defects valid for any doping level.
\section{\label{Disc}Results and Discussion}
\begin{figure}
\includegraphics[width=90mm]{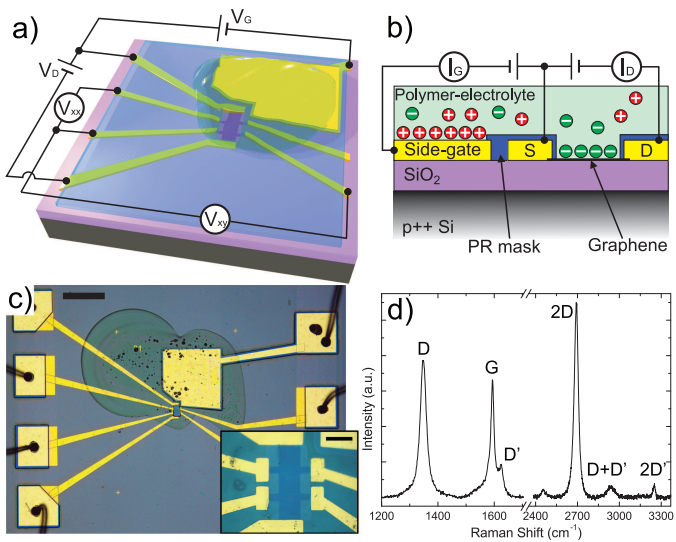}
\caption{a) Polymer electrolyte-gated graphene transistor and electrical configuration used for the measurements. b) Polymer electrolyte gating process. When biasing graphene with respect to the side-gate, Li$^+$(red) and ClO$_4^-$(green) ions migrate to form electric double layers near each electrode\cite{Das2008,Dhoot2006}. c) Optical micrograph of the device used in our experiments, scale bar 300$\mu$m. Inset: graphene channel; scale bar 30$\mu$m. d) Raman spectrum of defected SLG.}
\label{fig1}
\end{figure}

The defected samples are prepared as follows. A SLG film is grown by chemical vapor deposition (CVD) on a 25$\mu$m copper foil\cite{Bae2010,Bonaccorso2012Production}. It is then transferred on a Si+300nm SiO$_2$ substrate as described in Refs.\citenum{Reina2009,Bonaccorso2012Production}. The Hall bar geometry is defined by creating a photoresist (PR) mask by photo-lithography and removing the uncovered portion of the film by O$_2$/Ar reactive ion etching. A further PR layer is then spun on the sample and windows are opened by photolithography only on the graphene channel and side-gate area (see Fig.\ref{fig1}). The PR mask is then hard-baked at 145$^{\circ}$C for 5 minutes in order to improve its chemical stability\cite{MicrolithoBook}. The devices are then exposed to a mild O$_2$ inductively coupled plasma to introduce defects in the graphene channel\cite{Gokus2009}. The process is carried out at a pressure$\sim$150mTorr, with a power of 15W and for a few seconds (typically between 2 and 10s) depending on the desired defect concentration.

$E_F$ is moved by applying a droplet of polymer electrolyte, consisting of LiClO$_4$ and polyethylene oxide in the weight ratio 0.12:1, over both the device channel and the side-gate\cite{Das2008}. The working principle of this gating technique is shown schematically in Fig.\ref{fig1}(a,b). When a potential is applied between the side-gate and the graphene channel, free ions migrate and accumulate at the surface of the electrodes to form an electric double layer (EDL). Due to the large interfacial capacitance of the EDL, the compensation of these charges shifts $E_F$, much more than what can be achieved with standard dielectric gates\cite{GeimNatMat07, Pisana2007, Das2008}. In the case of standard dielectric-gating, the applied gate potential uniformly drops across the gate dielectric and it is therefore possible to directly estimate the induced charge through a capacitor model\cite{Novoselov04,Pisana2007}. In a polymer-electrolyte gated field-effect transistor, the applied gate potential drops across two nano-capacitors in series (one at the side-gate/electrolyte interface, the other at the electrolyte/channel interface), separated by an ionic conductive medium. In order to maximize the voltage drop across the channel/electrolyte interface, the side-gate/electrolyte capacitance must be the dominant one, hence the area of the gate electrode is significantly larger than that of the graphene channel, and a PR mask is fabricated on top of the device in order to minimize the direct contact area between metal electrodes and polymer electrolyte, thus reducing parasitic capacitance. Despite this, a voltage drop at the gate electrode cannot be excluded, and might lead to errors when correlating the applied gate voltage to the amount of induced charges in the graphene channel and $E_F$\cite{Xia2009}. To avoid possible systematic errors, $E_F$ in the graphene channel is directly evaluated by Hall-effect measurements. A 1$\mu$A direct current (DC) is applied between source and drain leads using a Keithley 2410 source-measurement unit, while longitudinal and transverse voltages (V$_{xx}$ and V$_{xy}$) are measured by means of a Keithley 2182A nanovoltmeter. The perpendicular magnetic field is applied using a permanent magnet with a surface field of 0.37T, as measured using a calibrated Gauss-meter. Another Keithley 2410 is used to apply the gate voltage, $V_g$. The overall performance of polymer-electrolyte gating is limited by the electrochemical stability of the polymer-electrolyte. When a $V_g$ higher than the electrochemical stability window of the polymer electrolyte is applied, electrochemical reactions, such as hydrolysis of residual water in the electrolyte\cite{azais2007,EfetovPRL}, can occur and permanently modify the graphene electrode, thus changing the total amount of defects. In order to avoid this, the maximum applied $V_g$ is$\sim\pm 2$V and the gate leakage current, a good indication of possible electrochemical reactions\cite{EfetovPRL,Ye2010}, is monitored and kept at$\sim$10$^{-10}$A. The absence of permanent modifications is confirmed by the repeatability of the measurements through several gate voltage sweeps.

Raman measurements are carried out at room temperature in a Renishaw InVia microspectrometer equipped with a 100$\times$ objective (numerical aperture 0.9). The spot size is$\sim$1$\mu\mathrm{m}$ and the incident power is kept well below 1mW in order to avoid heating effects. The excitation wavelengths are 514 and 633nm, chosen because these are the most commonly used for Raman characterization of graphene\cite{Ferrari2006, Cancado2011,Ferrari_NNano2013}.
\begin{figure}
\includegraphics[width=80mm]{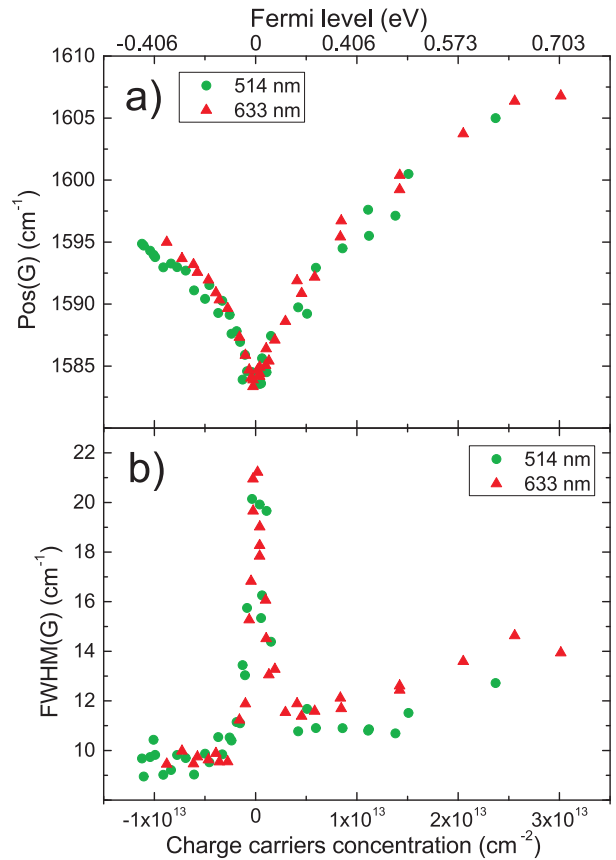}
\caption{Dependence of a) Pos(G) and b) FWHM(G) on (top x axis) $E_F$ and (bottom x axis) carrier concentration for excitation at 514 and 633nm}
\label{fig2p5a}
\end{figure}

\begin{figure}
\includegraphics[width=80mm]{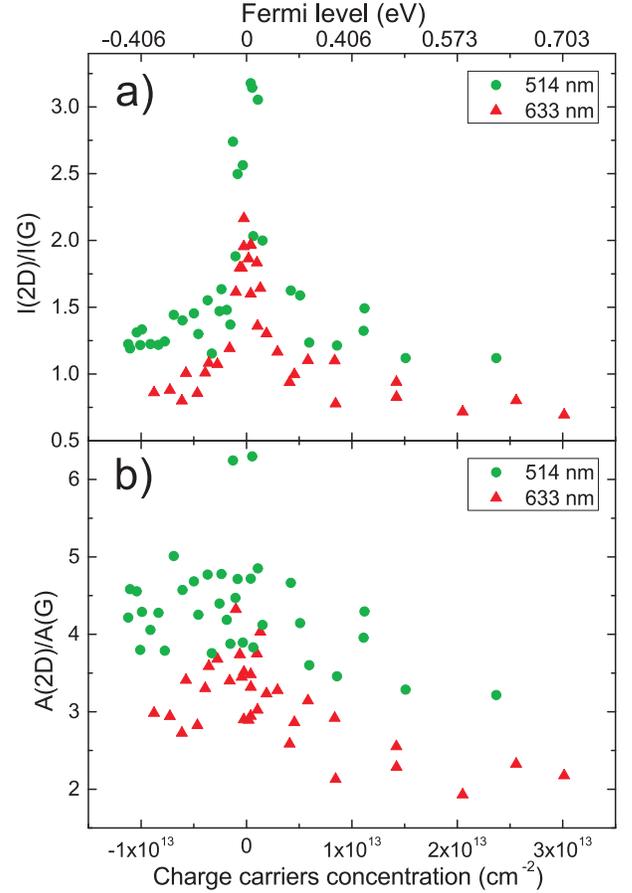}
\caption{Dependence of a)I(2D)/I(G) and b) A(2D)/A(G) on (top x axis) $E_F$ and (bottom x axis) carrier concentration for excitation at 514 and 633nm}
\label{fig2p5b}
\end{figure}

Fig.\ref{fig2p5a} plots Pos(G) and FWHM(G) as a function of $E_F$. As $E_F$ moves from the charge neutrality point, the G peak blue-shifts and narrows, consistent with what reported in pristine graphene in presence of moderate electrostatic doping\cite{Pisana2007, Das2008} with $E_F<<\hbar\omega_L/2$. The hardening of the G mode is due to nonadiabatic removal of a Kohn anomaly at $\Gamma$\cite{LazzeriPRL2006, Pisana2007}, and the reduction of width is due to Pauli exclusion principle inhibiting phonon decay into $e-h$ pairs when $E_F$ surpasses half the phonon energy\cite{Pisana2007}. Fig.\ref{fig2p5b} shows the doping dependence of the intensity and area ratio of 2D and G. Both decrease with increasing $E_F$ due to the effect of increased $e-e$ interaction\cite{Basko2009ee}. A(G) remains roughly constant with $E_F$\cite{Basko2009ee} (for $E_F<<\hbar\omega_L/2$), while I(G) is reduced close to the charge neutrality point due to the damping of the phonon decaying into $e-h$ pairs which increases FWHM(G). This gives a stronger doping dependence of I(2D)/I(G) than for A(2D)/A(G), Fig.\ref{fig2p5b}.

\begin{figure}
\includegraphics[width=80mm]{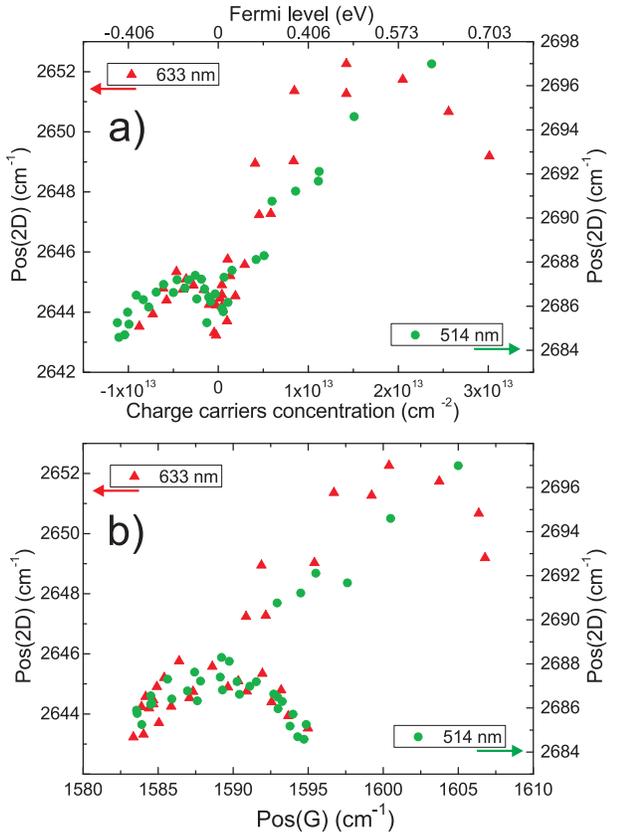}
\caption{a) Dependence of Pos(2D) on (top x axis) $E_F$ and (bottom x axis) carrier concentration for excitation at 514 and 633nm. b) Pos(2D) as a function of Pos(G).}
\label{fig2p5c}
\end{figure}

Fig.\ref{fig2p5c} plots the dependence of Pos(2D) on $E_F$. For $h-$doping Pos(2D) increases, while for $e-$doping, Pos(2D) slightly increases at first, then decreases as $E_F$ keeps rising\cite{Das2008}. This is due to a modification of the lattice parameters caused by doping, which changes the total number of charges, with a consequent stiffening/softening of the phonons\cite{Das2008}. The dependence of Pos(2D) on $E_F$ is different from that of Pos(G). Indeed, the latter always increases with $E_F$, as highlighted in Fig.\ref{fig2p5c}, where Pos(2D) is plotted against Pos(G). This allows to distinguish $e-$ from $h-$doping in graphene using Raman spectroscopy.
\begin{figure}
\includegraphics[width=90mm]{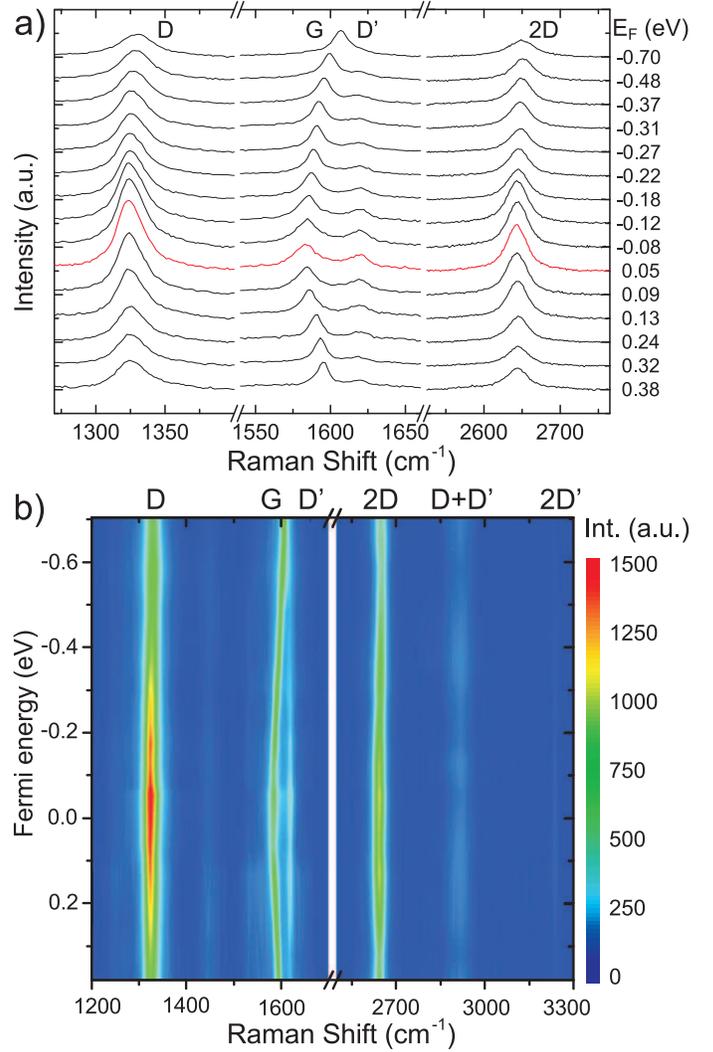}
\caption{a) Raman spectrum of defected graphene, measured at 633nm, for different $E_F$. The spectra are normalized with respect to I(G). The spectrum highlighted in red corresponds to the lowest $E_F$ probed in our experiments, separating $h-$doping (negative $E_F$) and $e-$doping (positive $E_F$). b) Contour plot of the same data, showing the intensity of the Raman signal as a function of $E_F$ and Raman shift.}
\label{fig2}
\end{figure}

\begin{figure*}
\includegraphics[width=160mm]{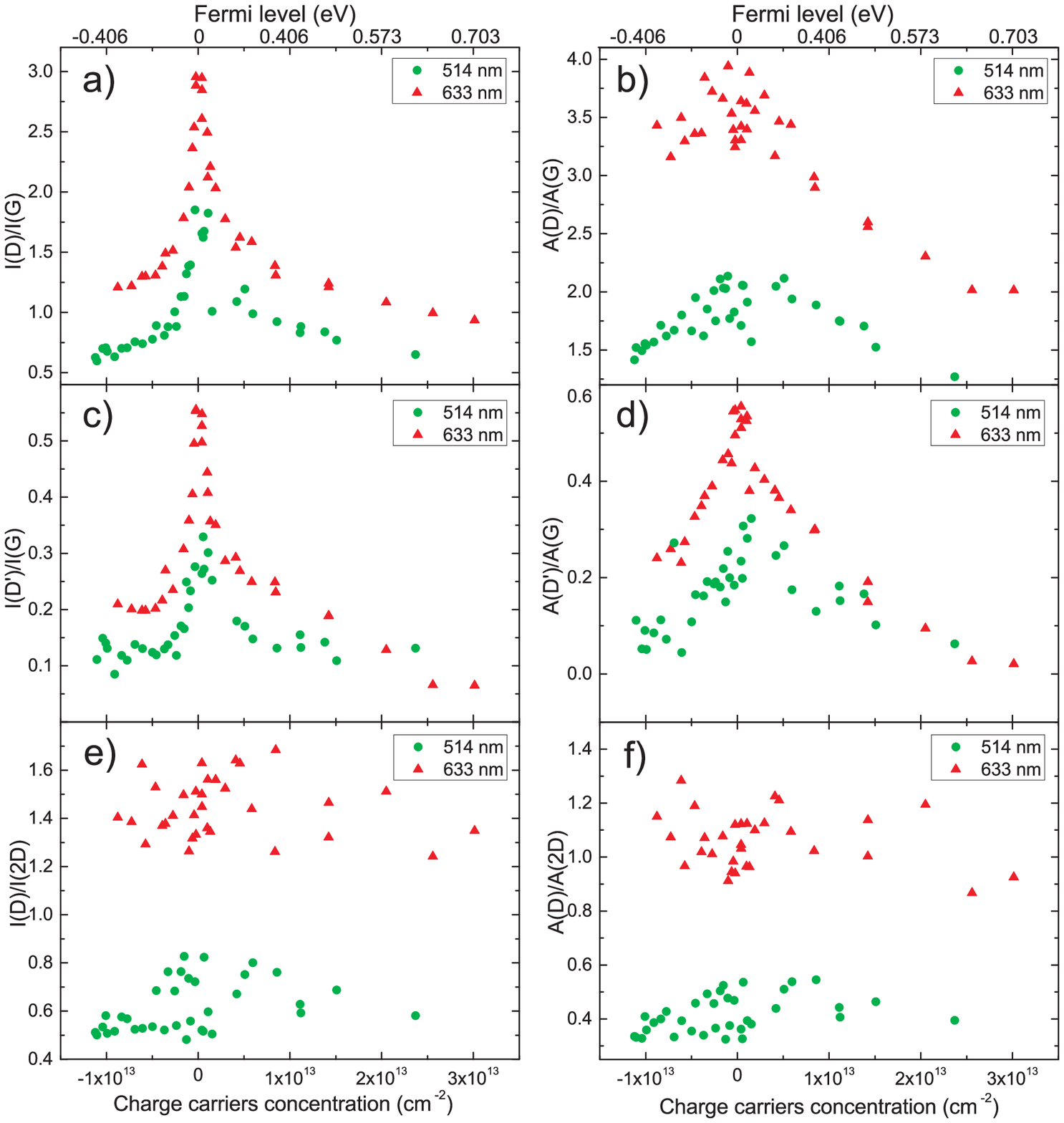}
\caption{a) I(D)/I(G), b) A(D)/A(G), c) I(D')/I(G), d) A(D')/A(G), e) I(D)/I(2D) f) A(D)/A(2D) as a function of (top $x$-axis) $E_F$ or (bottom $x$-axis) charge carrier concentration at two different excitation wavelengths for sample A, with $n^A_D\sim 4.4\times 10^{11}\,\mathrm{cm}^{-2}$.}
\label{fig3}
\end{figure*}

Now, we focus on the doping dependence of the main Raman signatures of defects in graphene: D and D'. We consider two samples, A, B. The defect concentration in is evaluated from I(D)/I(G) measured at low doping ($n<5\times 10^{11}\, \mathrm{cm}^{-2}$,$E_F$<90meV), with Eq.(\ref{eq14}). This gives $n^A_D\sim 4.4\times 10^{11}\, \mathrm{cm}^{-2}$ and $n^B_D\sim 2.5\times 10^{11}\, \mathrm{cm}^{-2}$. Fig.\ref{fig2} plots the Raman spectra of sample A as a function of doping. Figs.\ref{fig3}a,b) show that the evolution of the D peak with doping is similar to that of the 2D peak [compare Figs.\ref{fig3}a,b with Figs.\ref{fig2p5a}a,b], with a marked decrease in I(D)/I(G), for increasing $E_F$, Fig.\ref{fig3}a. The evolution of the peak's area ratio [Fig.\ref{fig3}b], more robust with respect to various perturbations of the phonon states than the height\cite{Basko2009ee}, shows a decrease with $E_F$. The same behaviour is observed for I(D')/I(G) and A(D')/A(G), Figs.\ref{fig3}c,d. Figs.\ref{fig3}e,f also indicate that I(D)/I(2D) and A(D)/A(2D) have no clear dependence on $E_F$, pointing to a similarity of the physical phenomena determining the doping-dependent Raman scattering for these two different Raman processes. Similar trends as in Fig.\ref{fig3} are also observed for the less defective sample B.
\begin{figure}
\centerline{\includegraphics[width=90mm]{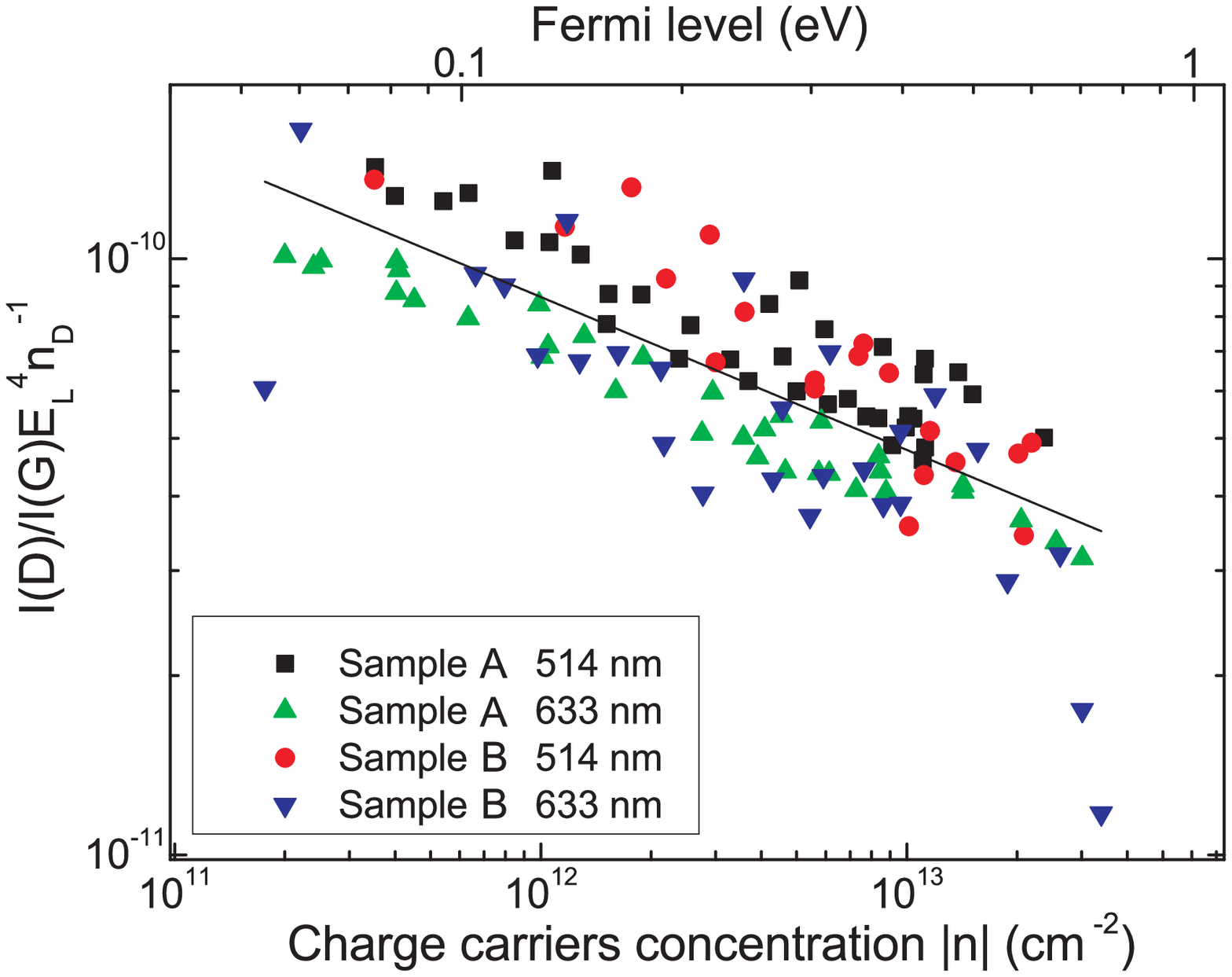}}
\caption{Doping dependence of I(D)/I(G) for both samples A and B re-scaled by the fourth power of $E_L$ and normalized by the amount of defects.}
\label{fig4}
\end{figure}

The decrease in I(D)/I(G) with $E_F$ can be described by a power relation. Re-scaling I(D)/I(G) by the fourth power of $E_L$ and normalizing by the amount of defects as measured at low doping, all the values collapse on the same line with slope $\alpha\sim-0.54$ in a logarithmic plot against the absolute value of $E_F$ (see Fig.\ref{fig4}).

We can thus modify Eq.(\ref{eq12}) for samples with non-negligible doping:
\begin{equation}\label{eq12m}
L_D^2\,{\rm(nm^{2})}=\frac{(1.2 \pm 0.3)\times10^{3}}{E_L^4(\mbox{eV}^4)}\left[\frac{\mathrm{I(D)}}{\mathrm{I(G)}}\right]^{-1}\{E_F[eV]\}^{-(0.54 \pm 0.04)}
\end{equation}
and Eq.(\ref{eq14}):
\begin{equation}
n_D[cm^{-2}]=(2.7 \pm 0.8) \times 10^{10} E_L^4[eV]\frac{I(D)}{I(G)}\{E_F[eV]\}^{0.54 \pm 0.04}
\label{nD2}
\end{equation}
Eqs.(3,\ref{nD2}) are valid for samples with a defect concentration corresponding to Stage 1, by far the most relevant for graphene production and applications, and for $E_F<E_L/2$, in order to avoid Pauli blocking effects on the intensity of peaks\cite{BaskoNJP, tandoping, Kalbac2010, Chen2011}. Combining I(D)/I(G) and FWHM(G) it is possible to discriminate between stages 1 or 2, since samples in stage 1 and 2 could have the same I(D)/I(G), but not the same FWHM(G), which is much larger in stage 2\cite{Ferrari2000,Cancado2011} Since most graphene samples in literature show doping levels$\sim$200-500meV and Raman characterization is mostly carried out with excitation wavelengths in the visible (1eV$>\hbar\omega_L/2>$1.5eV), Eqs.(\ref{eq12m},\ref{nD2}) cover the vast majority of experimental conditions in graphene science and technology\cite{Bonaccorso2012Production}.
\begin{figure}
\centerline{\includegraphics[width=80mm]{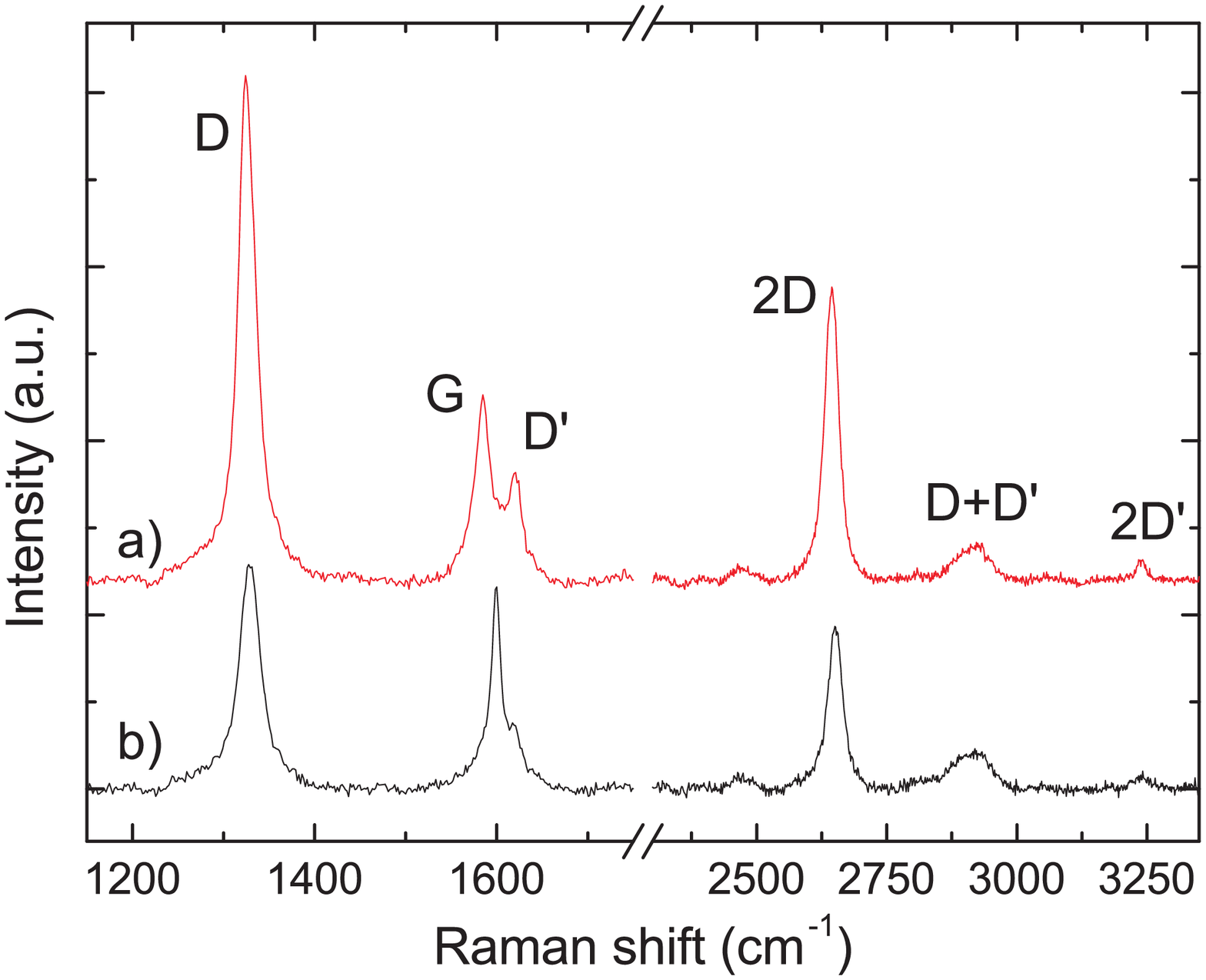}}
\caption{Raman spectra of sample A, acquired at 633nm, for a)$E_F\leq$100meV and b) $E_F\sim$500meV.}
\label{fig5}
\end{figure}

It is useful to consider a practical example on how Eqs.(\ref{eq12m},\ref{nD2}) may be used. Fig.\ref{fig5} plots a typical Raman spectrum of a defected doped sample (Fig.\ref{fig5}b) compared to the spectrum of the same sample, therefore with the same amount of defects, in the undoped case (Fig.\ref{fig5}a). In absence of external means to derive $E_F$, one could use Raman spectroscopy to evaluate $E_F$. Since Pos(G), FWHM(G), Pos(2D) and I(2D)/(G) for defected graphene within stage 1 evolve consistently with what reported for non defective samples\cite{Das2008}, one can use these to estimate $E_F\leq$100meV in case a) and $E_F\sim$500meV in case b). Comparing Pos(2D) with Pos(G) and using Fig.\ref{fig2p5c}b, it is possible to conclude that the sample is $h-$doped. Since I(D)/I(G)$\sim$1.24 at 633nm, if we ignore doping and use Eqs.(\ref{eq14},1) we get $n_D\approx1.3\times10^{11}$cm$^{-2}$ or $L_D\approx16$nm. Eqs.(\ref{nD2},3) instead give $n_D\approx 3.4\times10^{11}$cm$^{-2}$ or $L_D\approx10$nm. The defect density estimated taking into account doping is more than twice that from Eq.\ref{eq14}. This can make a difference for the optimization of methods of production and processing of graphene, especially for what concerns particular applications, such as transparent conductive films, where low sheet resistance may be achieved through high doping\cite{Bae2010}.
\begin{figure}
\centerline{\includegraphics[width=60mm]{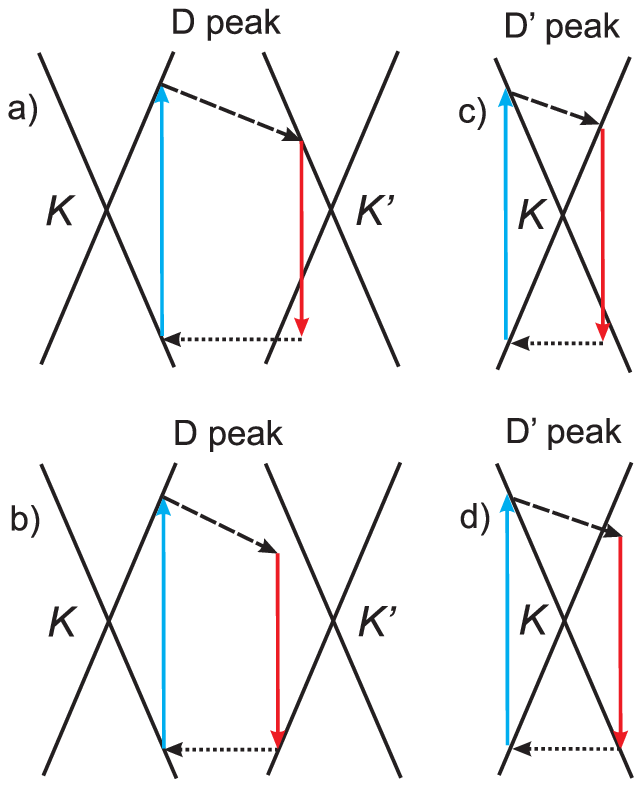}}
\caption{Raman processes giving rise to the D and D' peaks\cite{Ferrari_NNano2013}. a), b) Defect-assisted inter-valley one-phonon D peak. c), d) Defect-assisted intra-valley one-phonon D' peak. Vertical solid arrows represent inter-band transitions accompanied by photon absorption (upward arrows) or emission (downward arrows). Dashed arrows and horizontal dotted arrows represent phonon emission and elastic defect scattering.}
\label{fig4p5}
\end{figure}

We note that Ref.\citenum{nl2013} reported a doping dependence of I(D), which was found to sharply increase for $|E_F|\sim\hbar\omega_L/2$. This was assigned to quantum interference between different Raman pathways, similarly to the case of the G peak in Refs.\citenum{BaskoNJP,Kalbac2010,Chen2011}. However, Ref.\citenum{nl2013} suffers from poor accuracy in the determination of the charge carriers concentration, leading to results inconsistent with literature\cite{Das2008, Chen2011}. Indeed, in Ref.\citenum{nl2013} the graphene Raman peaks are found to evolve with doping as in Ref.\citenum{Das2008}, but at doping levels almost one order of magnitude higher. E.g., Pos(2D)$\approx2670cm^{-1}$ for $E_L=2.41eV$ is reached in Ref.\citenum{Das2008} for $n\approx 3.4\times 10^{13}cm^{-2}$ while in Ref.\citenum{nl2013} is observed for $n\approx 2.6\times 10^{14}cm^{-2}$. The inconsistency of Ref.\citenum{nl2013} is evident also in the analysis of the doping dependence of Pos(G). Ref.\citenum{Das2008} reported for pristine graphene $\delta Pos(G)/\delta E_F \approx 30 cm^{-1}eV^{-1}$ for $e-$doping and $\delta Pos(G)/\delta E_F \approx 42 cm^{-1}eV^{-1}$ for $h-$doping, as also confirmed by Ref.\citenum{Chen2011} and our present work, where we obtain $\delta Pos(G)/\delta E_F \approx 30 cm^{-1}eV^{-1}$ for $e-$doping and $\delta Pos(G)/\delta E_F \approx 39 cm^{-1}eV^{-1}$ for $h-$doping (see Fig.\ref{fig2p5a}). Ref.\citenum{nl2013} has a much weaker doping dependence of Pos(G), in particular $\delta Pos(G)/\delta E_F \approx 12 cm^{-1}eV^{-1}$ for $e-$doping and $\delta Pos(G)/\delta E_F \approx 19 cm^{-1}eV^{-1}$ for $h-$doping. This indicates that the actual doping level reached by Ref.\citenum{nl2013} is much smaller than what claimed and surely far from $|E_F|\approx\hbar\omega_L/2\geq 1.2eV$ necessary to achieve the blocking of Raman pathways\cite{BaskoNJP,Kalbac2010,Chen2011} for $\lambda$=514.5~nm (as used in Ref.\citenum{nl2013}), thus compromising the basis of their physical explanation for the observed increase of I(D).

Ref.\citenum{Casiraghi2013NanoRes} reported the I(D) dependence on back-gate bias in SLG, observing an increase in I(D)/I(G) with increasing gate voltage, in principle the opposite of what we report here in Fig.\ref{fig3}a. Ref.\citenum{Casiraghi2013NanoRes} attributed this to a change in the total amount of defects in graphene due to electrochemical reactions involving the water layer trapped at the interface between graphene and the silicon dioxide substrate. This is quite a different case with respect to that studied here, where the number of defects is kept constant as a function of doping, as confirmed by the repeatability of the Raman measurements through several gate voltage sweeps.

In order to understand the physical reason for the D peak decrease with $E_F$ reported in Fig.\ref{fig2}, we consider the Raman scattering processes in more detail. The G peak corresponds to the high frequency $E_{2g}$ phonon at $\Gamma$. The D peak is due to the breathing modes of six-atom rings and requires a defect for its activation\cite{TuinstraKoenig,Ferrari2000,Thomsen2000}. It comes from TO phonons around the BZ edge \textbf{K}\cite{TuinstraKoenig,Ferrari2000}, is active by double resonance (DR)\cite{Thomsen2000}, and is strongly dispersive with excitation energy\cite{pocsik}, due to a Kohn Anomaly at \textbf{K}\cite{Piscanec2004}. DR can also happen as intra-valley process, i.e. connecting two points belonging to the same cone around $\textbf{K}$ (or $\textbf{K}'$). This gives the D' peak. The 2D peak is the D peak overtone. The 2D' peak is the D' overtone. Since 2D and 2D' originate from a process where momentum conservation is satisfied by two phonons with opposite wavevectors, no defects are required for their activation, and are thus always present\cite{Ferrari_NNano2013,Basko2009ee}.

The Raman spectra can be modeled within second-order perturbation theory, by summing over all scattering pathways, expressed with the Fermi's golden rule to the fourth order\cite{Cardona_book}. Each of these pathways consists of the creation of an $e-h$ pair due to the interaction with incident excitation laser, one $e-ph$ and one electron-defect ($e-def$) scattering, and the $e-h$ pair recombination\cite{Venezuela2011,Ferrari_NNano2013}. The amplitude of each pathway is given by the corresponding matrix element, and by summing over all possible processes in the BZ, constructive and destructive interferences between the different quantum paths are fully taken into account. Both the photoexcited $e$ and $h$ may scatter, resulting in four different combinations, $ee$, $eh$, $he$, and $hh$. Taking into account the two relative orderings for the $e-ph$ and $e-def$ scattering, there are eight different contributions for each k-point in the BZ. E.g., the matrix element for a phonon-defect scattering pathway with the electron first scattering and emitting a phonon, and the hole then scattering with a defect, is given by\cite{Venezuela2011}:
\begin{widetext}
\begin{equation} \label{eq:ampl} K_{\mathrm{eh1}}^{\mathrm{pd}} = - \frac{\left\langle \mathbf{k}+\mathbf{q}, \pi \left| H_{e-em, \mathrm{out}}\right|\mathbf{k}+\mathbf{q}, \pi^* \right \rangle \left \langle \mathbf{k}\pi \left | H_D \right| \mathbf{k}+\mathbf{q}, \pi \right \rangle \left \langle \mathbf{k}+\mathbf{q}, \pi^* \left | \Delta H_{\mathbf{q} \nu} \right| \mathbf{k}\pi^* \right \rangle \left \langle \mathbf{k}\pi^* \left | H_{e-em, \mathrm{in}} \right | \mathbf{k}\pi  \right \rangle}{(E_L-E_{\mathbf{k}+\mathbf{q}}^{\pi^*} + E_{\mathbf{k}+\mathbf{q}}^{\pi} - \hbar \omega_{-\mathbf{q}}^{\nu}+2i \gamma_{\mathrm{tot}})(E_L-E_{\mathbf{k}+\mathbf{q}}^{\pi^*} + E_{\mathbf{k}}^{\pi} - \hbar \omega_{-\mathbf{q}}^{\nu}-2i \gamma_{\mathrm{tot}}  )(E_L-E_{\mathbf{k}}^{\pi^*} + E_{\mathbf{k}}^{\pi} -2i \gamma_{\mathrm{tot}})}.
\end{equation}
\end{widetext}

\begin{figure}
\centerline{\includegraphics[width=70mm]{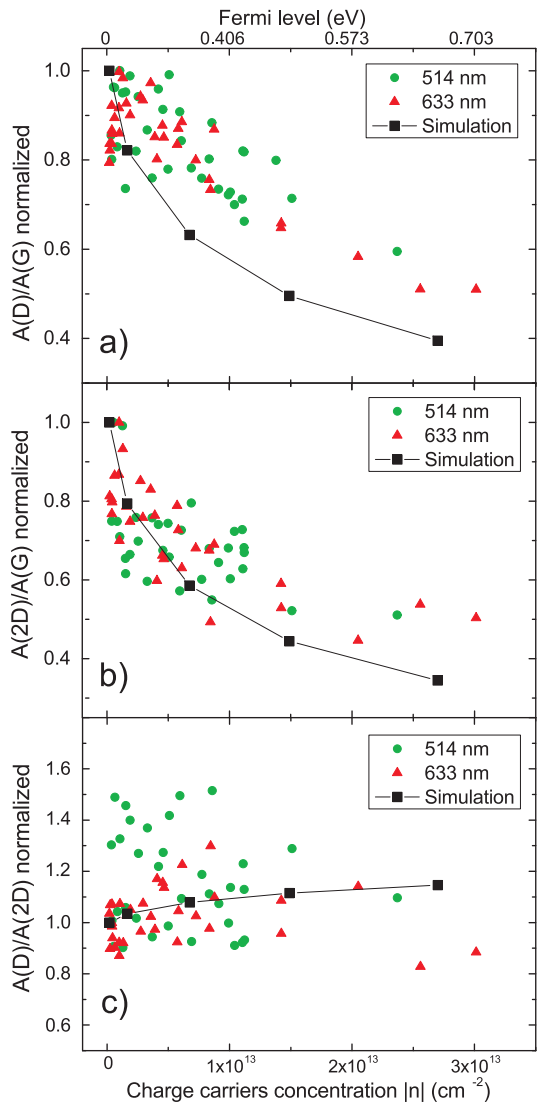}}
\caption{Evolution of a) A(D)/A(G), b) A(2D)/A(G), c) A(D)/A(2D) as a function of doping for sample A, normalized with respect to their values at low doping, compared to the calculated values. Green circles and red triangles show the experimental data at two different excitation wavelengths and the black squares indicate the calculated values. The line is to guide the eye.}
\label{fig7}
\end{figure}

where $E_{\mathbf{k}}^{\pi^{(*)}}$ is the energy of the electronic state in the valence (conduction) band at wave vector $\mathbf{k}$, $\omega_{\mathbf{q}}^{\nu}$ is the phonon frequency of branch $\nu$ at phonon momentum $\mathbf{q}$, and $\gamma_{\mathrm{tot}}$ describes the broadening of the intermediate electronic states. The nominator contains matrix elements corresponding to the light absorption and emission ($H_{e-em,\mathrm{in}}$ and $H_{e-em, \mathrm{out}}$, respectively), $e-ph$ scattering ($\Delta H_{\mathbf{q} \nu}$) and defect scattering ($\Delta H_{D}$). In the case of the 2D peak, the defect scattering matrix element is replaced by a second $e-ph$ scattering\cite{Venezuela2011,Ferrari_NNano2013}.

The Raman intensity is then obtained as:
\begin{equation}
\label{eq:ID} {I_D(\omega) \propto \sum_{\mathbf{q}, \nu} \left| \sum_{\mathrm{\mathbf{k}}, \alpha} K_{\alpha}^{pd}(\mathrm{\mathbf{k}}, \mathrm{\mathbf{q}}, \nu)   \right|^2 \delta(\omega-\omega_{\mathrm{\mathbf{q}}}^{\nu}), }
\end{equation}
and similarly for the 2D peak, but involving a summation over two phonon branches $\nu$ and $\mu$.

The contributions to the total broadening $\gamma_{\mathrm{tot}}$ due to $e-ph$, $\gamma_{\mathrm{el-ph}}$, and $e-def$ scattering, $\gamma_{\mathrm{D}}$, are determined by the corresponding scattering matrix elements (see Ref.\citenum{Venezuela2011}). In the presence of doping, a contribution due to $e-e$ scattering, $\gamma_{\mathrm{ee}}$, increasing with $E_F$, has to be included\cite{Basko2009ee}. In principle, these quantities depend on the wave vector and band index of the electronic state, thus they are inhomogeneous in the BZ. For simplicity, here, as in Ref.\citenum{Venezuela2011}, they are assumed to be independent of the electronic state.

Figs.\ref{fig2}(e,f) indicate a similar doping dependence of the D and 2D peaks. The A(2D) decrease with increasing doping is due to additional broadening of the intermediate $e-h$ states induced by $e-e$ interactions, with $\gamma_{\mathrm{ee}} \propto |E_F|$\cite{Basko2009ee}. Thus, the total broadening at a given $E_F$ is given by $\gamma_{\mathrm{tot}}=\gamma_{\mathrm{e-ph}}+\gamma_{D}+\gamma_{\mathrm{ee}}(E_F)$, where $\gamma_{\mathrm{ee}}(E_F) = 0.06|E_F|$\cite{Basko2009ee}. For the numerical simulation using Eq.~(\ref{eq:ID}), the band structure, $e-ph$ matrix elements, as well as defect scattering matrix elements are calculated using the fifth-nearest neighbor tight-binding model, as described in Ref.\citenum{Venezuela2011} on a 360$\times$360 grid of $\mathbf{\mathrm{k}}$-points. The phonon dispersion and the phonon eigenvectors needed for the calculation of the $e-ph$ matrix elements are calculated using standard density-functional perturbation theory as implemented in the Quantum ESPRESSO~\cite{QE} DFT package within local density approximation (LDA)\cite{Piscanec2004}. Norm-conserving pseudopotentials with a 55 Ry cutoff for the wave function are used, and the BZ is sampled by $32\times32$ $\mathbf{\mathrm{k}}$-points in the calculation of the electronic states. The phonons are calculated on a $8\times8$ $\mathbf{\mathrm{q}}$-point grid and for the Raman intensity they are interpolated into a  $120\times120$ grid. The $\delta$ functions in Eq.~(\ref{eq:ID}) are broadened into Lorentzians with FWHM$\sim$8cm$^{-1}$ to compensate for the finite computational grid\cite{Venezuela2011}. As defect, a weakened nearest-neighbor tight-binding hopping element is used, following Ref.\citenum{Venezuela2011}, with the perturbation being $\delta t$. The amount of defects is characterized by a parameter $\alpha_{\mathrm{hopp}}$ describing both defect density and magnitude of defect perturbation, $\alpha_{\mathrm{hopp}}=\delta t n_D$= 6.4$\times10^{13}$~eV$^2$cm$^{-2}$, as for Ref.\citenum{Venezuela2011}. A value of 68~meV for the doping-independent part of the broadening, consisting of contributions due to $e-ph$ and $e-def$ scattering, can well reproduce the experimentally observed trend of a decrease of the areas of the D and 2D peaks, as shown in Fig.\ref{fig7}. The $e-ph$ contribution in the present samples is estimated from our experimental data to be$\sim$31meV, following the procedure of Ref.\citenum{Basko2009ee}). Thus this the defect part is$\approx$40meV. To the best of our knowledge, this value has never been determined experimentally. The defect-related broadening depends on the number of defects and also the type of defects. Note that there is a difference of a factor of 4 between the definition of $\gamma$ of Ref.\citenum{Venezuela2011} and that used here, chosen to be consistent with the notation in Ref.\citenum{Basko2009ee}.

Fig.\ref{fig7} plots the doping dependence of A(D)/A(G), A(2D)/A(G) and A(D)/A(2D) for sample A, normalized with respect to their values at low doping. The area of the G peak is constant with doping for $\hbar\omega_L/2>E_F$\cite{Basko2009ee}, therefore the dependencies of A(D)/A(G) and A(2D)/A(G) with doping are representative of the A(D) and A(2D) trend, respectively. The agreement between experiment and theory is remarkable given the simple description of the doping in the simulation. The increase of the total broadening due to $\gamma_{\mathrm{ee}}$, as described in Ref.\citenum{Basko2009ee}, can well reproduce the trend of the decrease of the peaks' areas, strongly indicating that $e-e$ correlation is likely to be the most relevant cause for the observed experimental trends. As the concentration of charge carriers is increased, some of the other ingredients of the model, such as electron and phonon dispersions, might change. With doping, the graphene lattice parameter is expected to change, leading to a shift in the adiabatic energy of the phonon modes\cite{LazzeriPRL2006,Pisana2007}. This changes the position of the Raman peaks, but should not significantly alter their intensity. The $e-ph$ matrix elements are also expected to decrease slightly with increasing doping, $\sim$15\% for the doping range probed in this experiment\cite{AttaccaliteNL2010}, thus representing a correction to our analysis. More detailed studies are required to fully understand the interplay between the different effects.
\section{Conclusions}
We studied the dependence of Raman spectrum of defected graphene on the charge carrier concentration, by combining polymer electrolyte gating with \textit{in situ} Hall-effect and Raman measurements. For a given number of defects, the intensities and areas of the D and D' peaks decrease with increasing doping. Considering all scattering processes within the DR framework, we interpret the doping-induced intensity variation as due to an increased total scattering rate of the photoexcited electrons and holes, resulting from the doping-dependent strength of electron-electron scattering. This analysis paves the way for the experimental evaluation of the different sources of broadening in the electronic states in graphene and a better understanding of the role and type of defects on its physical properties. This study highlights the importance of taking into account the doping level when determining the amount and the type of defects from the intensity of the D-peak. We therefore presented general relations between D peak intensity and defects valid for any doping level.
\begin{acknowledgements}
We thank D. M. Basko and N. Bonini, A. Uppstu, N. Marzari, F. Giustino, M. Lazzeri, A. S. Dhoot for useful discussions. We acknowledge funding from EU projects GENIUS, CARERAMM, MEM4WIN, EU Graphene Flagship (contract no. 604391), ERC grant Hetero2D, a Royal Society Wolfson Research Merit Award, EPSRC grants EP/K01711X/1, EP/K017144/1, EP/L016087/1
\end{acknowledgements}

\end{document}